\documentclass[conference]{IEEEtran}
\usepackage[utf8]{inputenc}
\usepackage{caption}
\usepackage{url}

\usepackage{xcolor}
\title{GraphAide: Advanced Graph-Assisted Query and Reasoning System}
\author{
    \IEEEauthorblockN{
       Sumit Purohit, George Chin, Patrick S Mackey, Joseph A Cottam
    }
    \IEEEauthorblockA{
        Pacific Northwest National Laboratory \\
        Richland, Washington 99352 \\
        \{sumit.purohit, george.chin, patrick.mackey, joseph.cottam\}@pnnl.gov
    }
}

\usepackage{natbib}
\usepackage{graphicx}

\begin{document}

\maketitle
\begin{abstract}
    
Curating knowledge from multiple siloed sources that contain both structured and unstructured data is a major challenge in many real-world applications. Pattern matching and querying represent fundamental tasks in modern data analytics that leverage this curated knowledge. The development of such applications necessitates overcoming several research challenges, including data extraction, named entity recognition, data modeling, and designing query interfaces. Moreover, the explainability of these functionalities is critical for their broader adoption. 

The emergence of Large Language Models (LLMs) has accelerated the development lifecycle of new capabilities. Nonetheless, there is an ongoing need for domain-specific tools tailored to user activities. The creation of digital assistants has gained considerable traction in recent years, with LLMs offering a promising avenue to develop such assistants utilizing domain-specific knowledge and assumptions. 

In this context, we introduce an advanced query and reasoning system, \textit{GraphAide}, which constructs a knowledge graph (KG) from diverse sources and allows to query and reason over the resulting KG. \textit{GraphAide} harnesses both the KG and LLMs to rapidly develop domain-specific digital assistants. It integrates design patterns from retrieval augmented generation (RAG) and the semantic web to create an agentic LLM application. \textit{GraphAide} underscores the potential for streamlined and efficient development of specialized digital assistants, thereby enhancing their applicability across various domains.
\end{abstract}

\section{Introduction}
Large Language Models (LLMs) \cite{brown2020language} represent the cutting edge of generative artificial intelligence (GenAI) and machine learning research and development. They have been adopted at an extraordinary rate across a range of disciplines, including scientific research, engineering, economics, and social sciences. By enabling domain experts to utilize pre-trained, ready-to-use models, LLMs have democratized the application development process, catalyzing innovations in both scientific and technological domains. The development of LLM-based applications remains a dynamic area of research, offering superior performance in tasks such as summarization, correlation, and inference across various input sources \cite{zhang2024benchmarking}.

Knowledge Graphs (KGs) are formal representations of key concepts as entities and the relationships between them \cite{berners2023semantic, bizer2023linked}. KGs serve as powerful tools that can be employed in diverse user environments as foundational reference knowledge sources. However, constructing these KGs presents substantial challenges due to the scale of input data, the heterogeneity of use-case-specific concepts, and the costs associated with KG construction. LLMs have demonstrated impressive accuracy in extracting relevant entities and forming relationships, significantly contributing to KG construction.

A major challenge in developing accurate and consistent capabilities based on LLMs is \textit{hallucination} \cite{yao2023llm}, which occurs when LLMs generate non-existent facts in response to user queries. The memorization of training data by LLMs and the subsequent reliance on corpus-based heuristics are significant factors contributing to hallucinations \cite{mckenna2023sources}. Addressing this issue is crucial for the reliability and trustworthiness of LLM-derived applications.

Retrieval-augmented generation (RAG) \cite{lewis2020retrieval} is a well-established design pattern aimed at mitigating hallucination in Large Language Models (LLMs) by providing additional context during the generation of responses to user queries. This additional context is domain-specific and acts as \textit{grounding} for the model's generative capabilities. A foundational RAG-based system employs a vector database to store \textit{embeddings} of a domain-specific knowledge corpus and utilizes semantic similarity algorithms to retrieve the \textit{context} relevant to the user query. This context is then combined with the user query and forwarded to the LLM, serving as a guardrail for its generation process. The RAG-based approach has demonstrated efficacy in reducing hallucination by constraining the LLM's generative output to the localized region of the provided context.

However, a vector search-based method is constrained by the semantic similarity between the query and the corpus. It fails to leverage the structural context, such as the relationships between documents within the corpus, their metadata, or the associated reasoning. Knowledge Graphs (KGs) offer a robust solution to these limitations by efficiently storing domain-specific information and establishing relationships between information sources dispersed across documents in the corpus. KGs can also significantly enhance various components of an LLM-based application, including retrieval, summarization, and inference.

By integrating KGs into the design pattern, the system can utilize both the semantic and structural context, thereby improving the accuracy and consistency of the generated responses. This comprehensive approach enables a more nuanced understanding and utilization of the knowledge corpus, offering a sophisticated strategy to further reduce hallucination and enhance the application’s overall performance.

\section{Related Work}\label{sec:relwork}
The integration of generative artificial intelligence (GenAI) and semantic technologies is an active and burgeoning area of research. Existing Knowledge Graph (KG)-based LLM capabilities often focus on specific applications. A baseline RAG approach \cite{gao2023retrieval} transforms unstructured data into multiple chunks, creates embeddings for these chunks, and stores them in a vector database. These embeddings are then used to perform semantic similarity matching for an input user query, retrieving the most similar chunks. This retrieval process constructs the \textit{context} to augment user queries sent to the LLM. The RAG approach has shown promising results in addressing challenges such as hallucination, outdated knowledge, and the explainability of generated responses. However, the baseline RAG is limited by its reliance on semantic similarity.

GraphRAG \cite{edge2024local} advances this field by presenting a query-focused summarization task (QFS) that builds efficient text indices over a KG. GraphRAG leverages community detection to partition the graph index and uses it to generate "global answers" to user queries. Other approaches, such as the work by Trajanoska et al. \cite{trajanoska2023enhancing}, focus on creating KGs from sustainability-related text and evaluating their quality with specialized models for joint entity and relation extraction. Similarly, Ban et al. \cite{ban2023query} introduce a set of prompts to extract causal graphs for causal structural learning. A common challenge with these approaches is their limited extensibility, as they do not leverage state-of-the-art open-source application development frameworks. There exists a significant opportunity to utilize KGs more effectively in various phases of LLM-based application development and usage.

We introduce GraphAide, an advanced capability based on Large Language Models (LLMs) that offers insights into domain-specific data and allows users to ask natural language questions. GraphAide presents a methodology and reference architecture to develop advanced GenAI applications using different components such as LLMs, vector database, and graph database. GraphAide employs a modular, extensible, and user-friendly retrieval-augmented generation (RAG) approach, incorporating both vector and graph databases. The novelty of GraphAide lies in its comprehensive design for integrating GenAI and semantic web technologies. GraphAide curates multi-source data and constructs ontology-guided knowledge graphs to address the challenges faced by baseline LLM-only applications. The platform also provides efficient data persistence, enabling interactions via a natural language question-answering interface.

GraphAide combines a robust graph matching approach with the semantic similarity provided by vector databases to enhance the accuracy and explainability of LLM-generated responses. Developed as an extensible agentic RAG application, GraphAide reuses data ingestion, entity disambiguation, storage, query, and reasoning components throughout the application lifecycle. Key capabilities of GraphAide include: \begin{itemize}
\item \textbf{Ontology-guided Knowledge Graph construction:} Handles multi-value, multi-source, temporal attributes from common data formats such as .txt, .html, .pdf, etc. 
\item \textbf{Generation and maintenance of embeddings:} Configures vector databases for storage and retrieval, maintaining multiple vector databases and dynamically selecting sources based on user queries.
\item \textbf{Template-based ontology configuration:} Enables dynamic input ontology configuration and generation of multiple KG versions for a given source dataset. 
\item \textbf{Enhanced context generation:} Combines vector-based retrieval with graph-based subgraph matching to improve the relevancy of LLM-generated responses.
\item \textbf{External function execution capability:} Enriches context by retrieving multi-modal data sources, such as temporal data from time-series databases. \end{itemize}

GraphAide represents a significant advancement in the development of domain-specific digital assistants, integrating state-of-the-art GenAI and semantic technologies for improved performance and extensibility.

\section{Architecture}\label{sec:arch}
GraphAide is an agentic retrieval-augmented generation (RAG)-based system that employs multiple LLM instances to perform diverse tasks. \textit{Agent}-based LLM applications utilize the reasoning capabilities of an LLM to determine the appropriate sequence of actions in response to user queries. Unlike hardcoded instructions (i.e., \textit{chains}), an \textit{agentic application} can comprehend the LLM's response and construct subsequent queries dynamically. GraphAide is a hybrid application leveraging both chains and agents to handle a variety of user queries. GraphAide's hybrid approach, utilizing both chains and agents, enables the system to effectively manage a wide array of user queries, ensuring accurate and contextually relevant responses. 

\begin{figure*}[ht]
    \centering
    \includegraphics[width=.99\textwidth, height=.5\textheight]{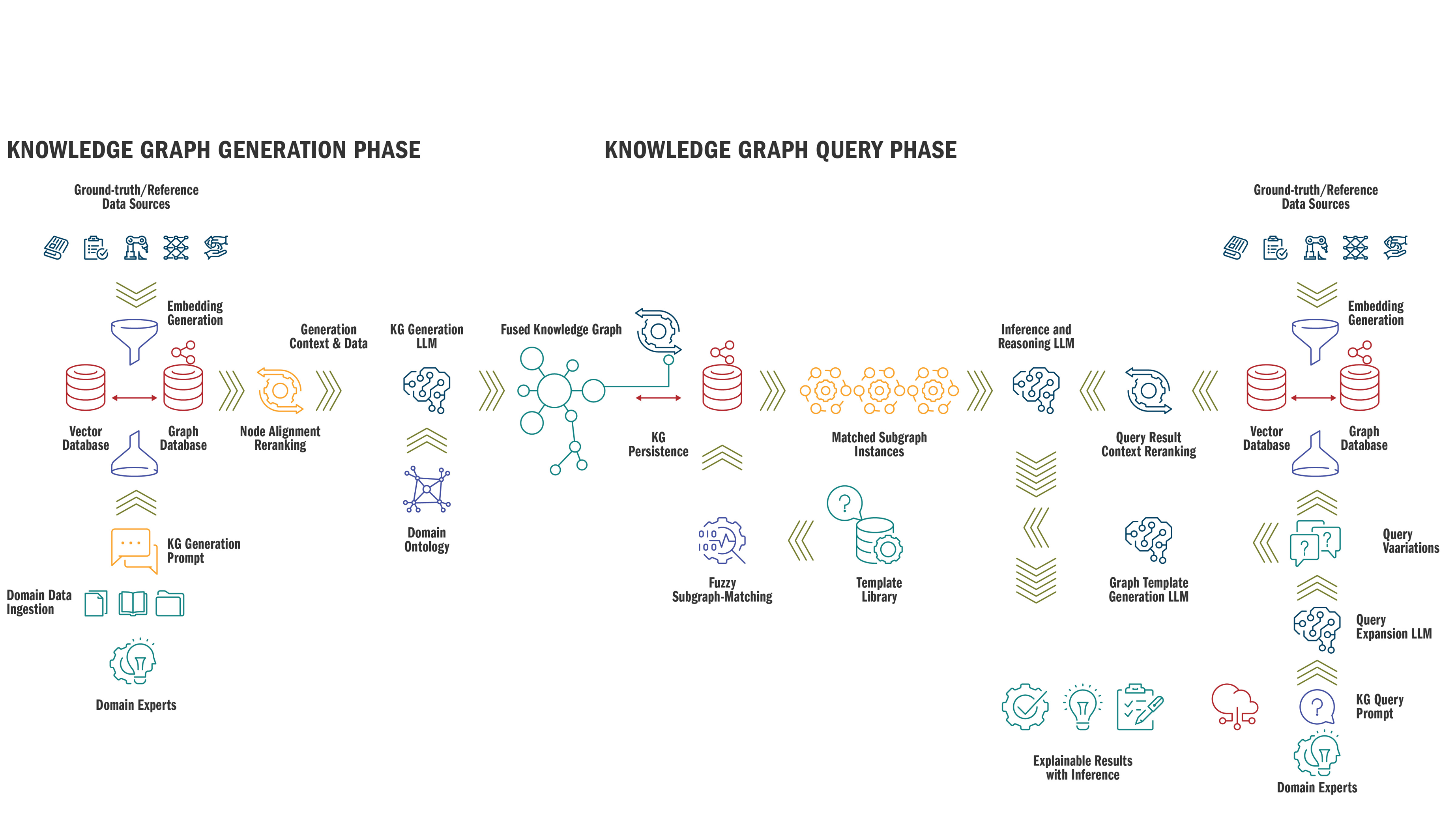}
    \caption{GraphAide Architecture, representing different phases and agentic workflow}
    \label{fig:graphaid_arch}
\end{figure*}

GraphAide delineates specific \textit{phases} corresponding to the set of features required to complete a user task. As illustrated in Figure \ref{fig:graphaid_arch}, the \textit{curation} phase combines information from multiple sources to construct a Knowledge Graph (KG), which serves as a knowledge repository alongside the vector database. In the \textit{exploration} phase, GraphAide provides a user interface that allows querying the knowledge sources and generating responses in a user-specified format, accompanied by explanations and reasoning.

GraphAide decomposes the query and reasoning research problem into following sub-problems and addresses them using a graph-based RAG approach:
\begin{itemize}
    \item Data ETL and Normalization
    \item Named-Entity Recognition
    \item Domain-aware semantic annotation
    \item Graph Modeling and Storage
    \item Query interpretation and context generation
    \item Query expansion and candidate generation
    \item Summarization and explainable response generation
\end{itemize}

\subsection{Data Ingestion}\label{sec:dataingest}
Domain-specific datasets are available in multiple structured and unstructured formats, depending on the use cases. GraphAide creates an abstraction layer that allows end users to work with commonly used data formats such as plain text, CSV, PDF, etc. A collection of source data (i.e., files) is ingested into GraphAide, followed by a process of text splitting and chunking to support efficient search over vector data and to manage data larger than the maximum context window size (in number of tokens) of an LLM. GraphAide employs a byte pair encoding (BPE)-based tokenizer \cite{gage1994new} that segments input text into sub-word units, enabling the system to handle a wide variety of languages. The BPE-based encoding is also supported by OpenAI embedding models \cite{radford2019language} and has been shown to perform well across different query scenarios due to its lossless, reversible, and efficient encoding. Additionally, GraphAide uses the $RecursiveCharacterTextSplitter$, which recursively examines characters in the input data to dynamically identify the optimal splitting strategy. GraphAide employs static $chunk\_size$ and $chunk\_overlap$ values, as suggested in related work \cite{karpukhin2020dense}, and also allows users to configure the text splitter for specific application domains.

\subsection{Data Storage}\label{sec:datastorage}
Persistence of knowledge curated from multiple sources is critical for the reproducibility and incremental updates of the underlying digital assistant. In contrast to  basic RAG, GraphAide persists contextual knowledge in both vector and graph databases. After splitting and chunking, as discussed in Section \ref{sec:dataingest}, the source data is converted into a collection of $documents$, where each $document$ serves as an atomic knowledge source used to answer user queries. In a RAG-based LLM application, \textit{embedding}-based semantic similarity is a key feature. Embeddings are numerical representations of information, where semantically similar content is placed closer together in the embedding space. GraphAide uses a vector database to store embeddings of domain-specific reference datasets or general-purpose knowledge bases, such as WikiData \cite{vrandevcic2014wikidata}. GraphAide provides a simple API to ingest datasets into the vector database and persist them locally for future use. The novelty of GraphAide lies in its use of a graph database to store additional structural information. GraphAide's API offers domain-specific graph modeling and graph persistence functionality. The graph database is used in subsequent phases to enhance the performance of retrieval operations.

\subsection{Knowledge Graph Generation}\label{sec:kggen}
Generation of an ontology-guided Knowledge Graph (KG) from heterogeneous sources is one of the key contributions of GraphAide. GraphAide defines a KG generation phase that supports operations such as named entity extraction, relationship extraction, entity disambiguation, KG encoding, and KG maintenance. It employs a prompt-based, few-shot approach to instruct an LLM to create a knowledge graph. GraphAide supports the dynamic selection of an ontology or a list of node and edge types as input to the generation phase, which restricts the LLM's response schema, ensuring it generates an attributed subgraph for a given input text. For the KG generation phase, an example query could be a paragraph from an unstructured source, such as a news article. GraphAide uses Pydantic \cite{pydantic2}, an annotation-based schema validator, to maintain the consistency of the generated subgraph throughout the entire KG generation phase.

\subsection{Subgraph Query Generation}\label{sec:subquery}
GraphAide's \textit{query phase} begins with a natural language user query, as shown in Figure \ref{fig:graphaid_arch}. User queries are often incomplete and imprecise, which can lead to reduced retrieval accuracy and relevance when using vector databases. To address query-document mismatch in sparse retrieval, LLM-based query expansion has been shown to alleviate these challenges \cite{ayoub2024case}. GraphAide applies a similar approach to embedding-based dense retrieval, using a template-based method to generate additional versions of the input query while preserving the key entities and relationships mentioned in the original query. Subsequently, GraphAide leverages LLMs to convert the expanded set of queries into attributed subgraph matching templates, as illustrated in Figure \ref{fig:graphaid_arch}. By combining expressive subgraph matching with vector-based semantic similarity, GraphAide enhances retrieval accuracy. The GraphAide API supports commonly used graph databases, such as Neo4J, and can auto-generate Cypher queries for the expanded query set. Moreover, the GraphAide API provides different prompt templates to generate queries for domain-specific graph data models based on custom schema.

\subsection{Explainable Result Generation}
GraphAide simplifies the execution of the templates generated in section \ref{sec:subquery} for a given graph database and facilitates subgraph matching. It offers an API to combine subgraph matching results with vector-based semantic similarity, creating a local context for the LLM during the query phase. Unlike the KG-generation phase, the query phase uses templates to summarize and explain results from multiple retrieval sources, rather than focusing on entity disambiguation. GraphAide executes multiple parallel requests to the LLM for the expanded query set and makes a secondary call to aggregate all the responses, generating a single coherent response for the user query. Auto-generation of complex templates with multiple constraints and clauses is a part of our future work. We plan to use a few-shot prompt approach to enhance the template generation process.


\section{Experiment}\label{sec:exp}
For our experiment, we instantiated GraphAide using OpenAI LLM, Chroma vector database, Neo4j graph database, and LangChain application development framework. We used GPT-4o model that has larger context window and improved reasoning and deduction capabilities. GPT-4o has also shown improved multi-lingual support and that makes it relevant for our use case discussed below. We leveraged the work of two existing DARPA efforts. Active Interpretation of Disparate Alternatives (AIDA) \cite{darpa_aida} developed multi-hypothesis semantic engines to generates and understand different interpretations of events, situations, and trends from a variety of unstructured sources. Generating Alternative Interpretations for Analysis (GAIA) \cite{zhang2018gaia} developed automated solutions to extract entities, emerging events, situations, and trends of interest for Ukraine-Russia political conflict scenario. An ontology \cite{aida_interchange_format} is also constructed containing approximately 190 entity types, 140 event types, and 50 relation types. A DARPA version of WikiData (DWD) is also developed as a reference knowledge base for the entity disambiguation process.
\begin{figure*}[ht]
    \centering
    \includegraphics[width=.99\textwidth, height=.4\textheight]{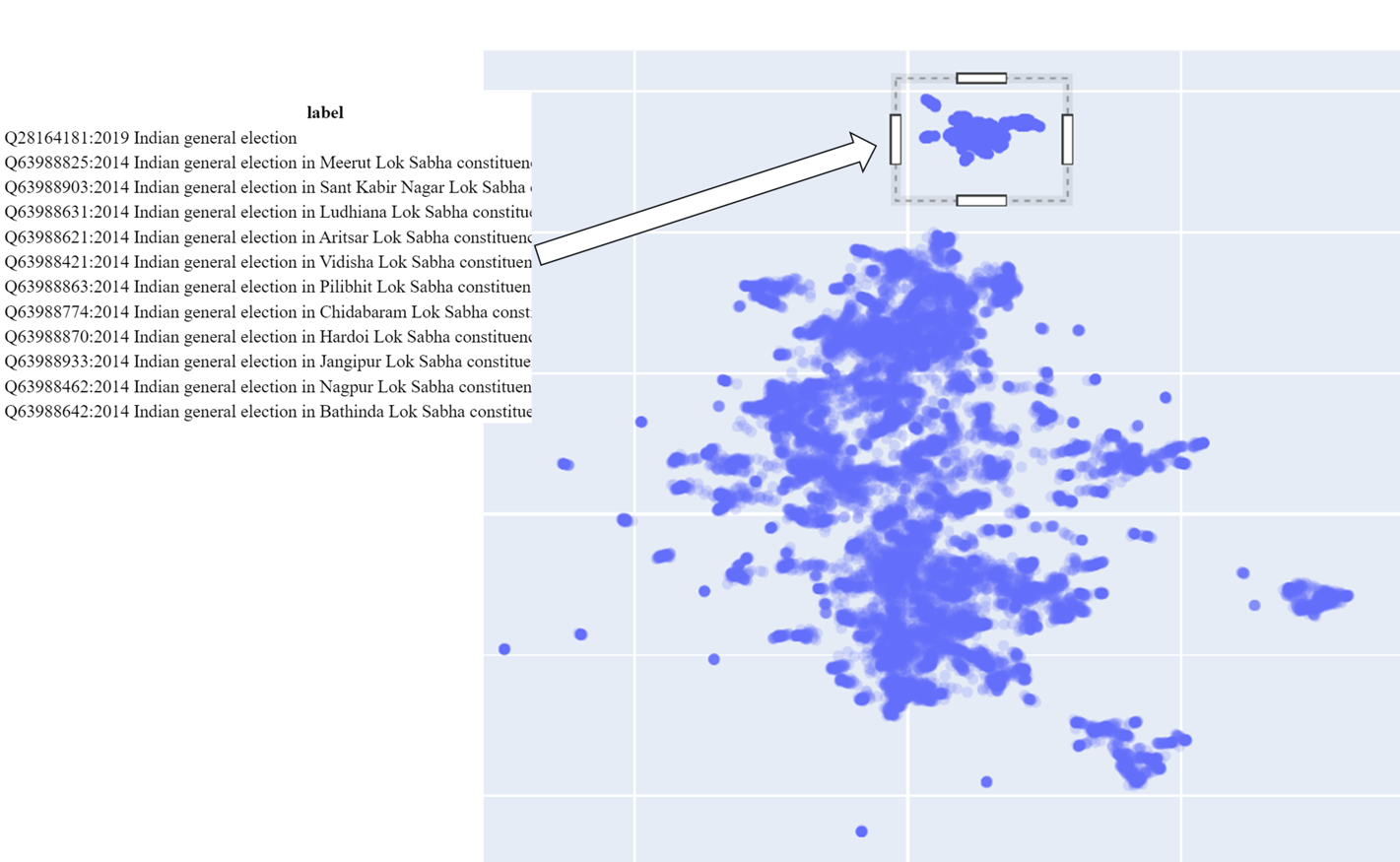}
    \caption{DARPA WikiData (DWD) Embeddings Used in Retrieval}
    \label{fig:dwd_emb}
\end{figure*}
To demonstrate GraphAide's capabilities, we used AIDA source data in text format and a Web Ontology Language (OWL) ontology in Terse RDF Triple Language (TURTLE) format to generate a KG. We evaluated the quality of the generated KG by comparing it with a GAIA-generated KG. We also used DWD as a reference knowledge base, stored in Chroma vector database for retrieval during KG generation phase. We curated natural language description of DWD identifiers (i.e., QIDs) and constructed \textit{documents} to store in the vector database. We use QIDs as the document identifier and use LLMs to disambiguate mentions to top-ranked QID. GraphAide also  generate UMAP embeddings of vector database to visualize the clusters of topics discussed in the KG, as shown in Figure \ref{fig:dwd_emb}. 
\begin{figure*}[ht]
    \centering
    \includegraphics[width=.99\textwidth, height=.4\textheight]{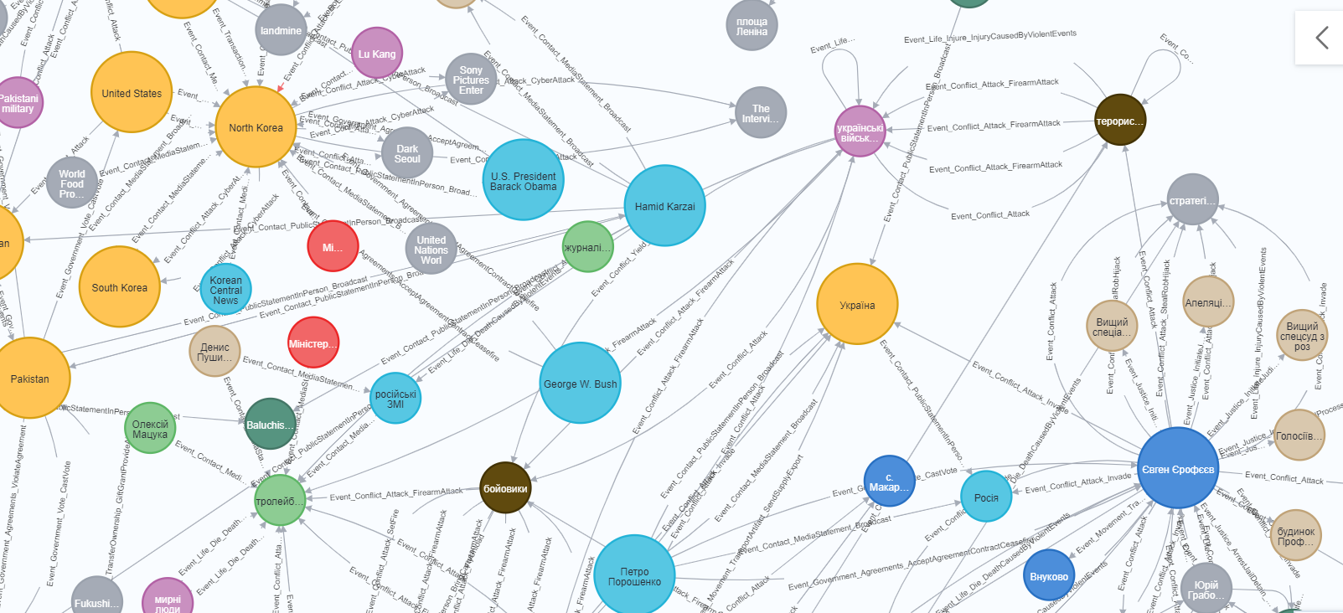}
    \caption{GraphAide generated KG from News Sources}
    \label{fig:ukr14kg}
\end{figure*}

GraphAide's \textit{curation} phase processed 1,846 news articles and generated a KG, as shown in Figure \ref{fig:ukr14kg}. An ontology-guided and WikiData-based disambiguation agent extracts entity mentions from the text and performs RAG-based node re-ranking to identify the closest matched QID. GraphAide iteratively performs text chunking, generates partial KGs from the given chunks, and loads them into the Neo4j graph database. 

\begin{figure*}[ht]
    \centering
    \includegraphics[width=.99\textwidth, height=.3\textheight]{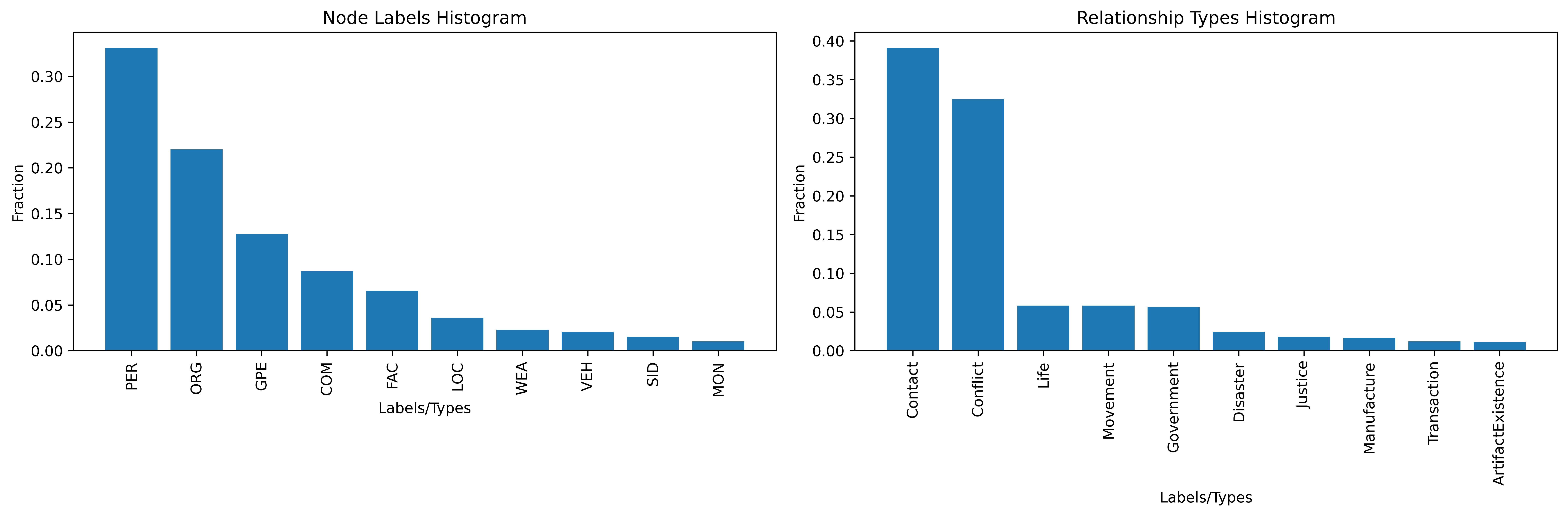}
    \caption{Type Distributions (Top-10 ) of GraphAide generated KG}
    \label{fig:ukr14distri_graphaide}
\end{figure*}

\begin{figure*}[ht]
    \centering
    \includegraphics[width=.99\textwidth, height=.3\textheight]{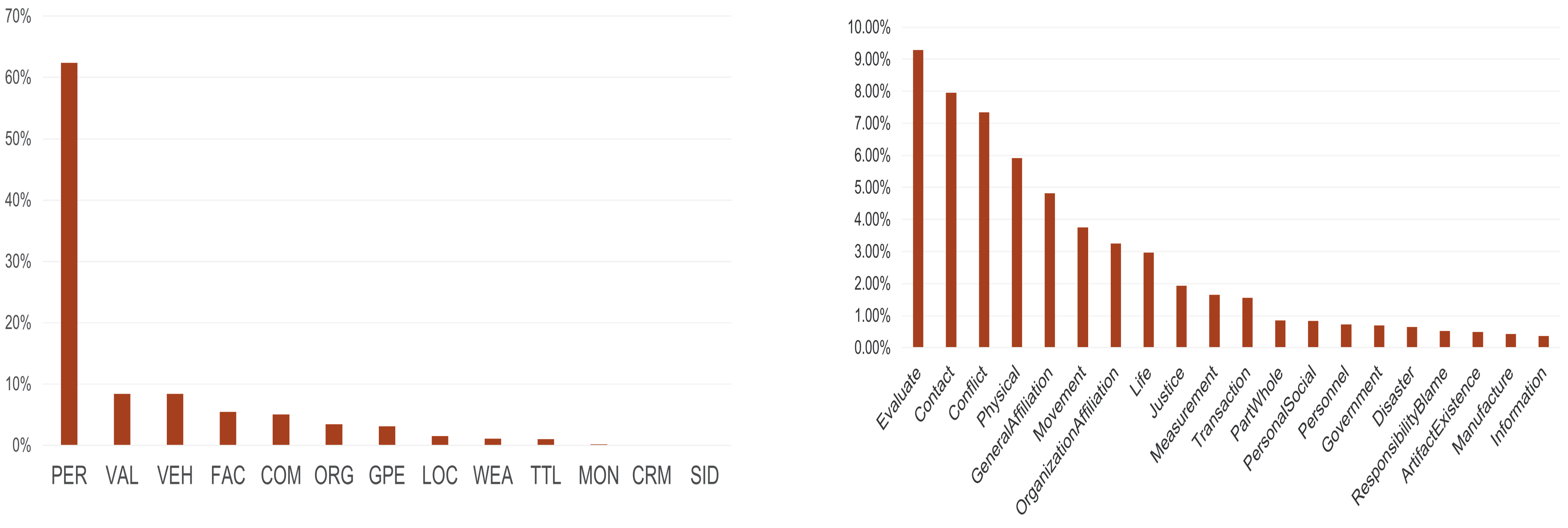}
    \caption{Type Distributions of Baseline KG. Node (left), Relationship (right)}
    \label{fig:ukr14distri_baseline}
\end{figure*}

In contrast to the basic RAG, GraphAide's KG-based hybrid-RAG approach provides higher specificity and cross-document inference to construct a richer context for LLM queries. Additionally, GraphAide's modular design allows dynamic selection from multiple retrieval sources for benchmarking purposes. We performed a qualitative evaluation of the generated responses in each phase. For the generation phase, LLM query templates were designed to extract natural language mentions from the input text, perform entity disambiguation using DWD, and perform type assignment (nodes and edges) based on the input ontology. As shown in Figure \ref{fig:ukr14kg}, the KG schema is constrained by the domain-specific ontology. The resulting KG has demonstrated better NER and relation extraction quality than the baseline KG \cite{zhang2018gaia, purohit2021semantic}. An extracted node type follows a hierarchical schema to represent the most specific concept about the node. For example, \(GPE\_UrbanArea\_City\) represents a class \(City\), which is a subclass of an urban area, itself a subclass of a geopolitical entity. Similarly, \(ConflictAttack\_FirearmAttack\) represents an attack event related to some conflict that involves firearms. As shown in Figure \ref{fig:ukr14distri_baseline}, the baseline KG shows a node type imbalance, where the most frequently extracted type is ``PER'' and the rest of the node types constitute a small fraction of semantic annotations. Similarly, non-binary relationship types such as ``Evaluate'' and ``GeneralAffiliation'' dominate the relationship extraction. In contrast, GraphAide extracts a more diverse and dense node type distribution, as shown in Figure \ref{fig:ukr14distri_graphaide}. Similarly, for the edge types, GraphAide performs better in extracting event-based types, given the input data describes many temporal events.

\begin{figure*}[ht]
    \centering
    \includegraphics[width=.9\textwidth, height=.45\textheight]{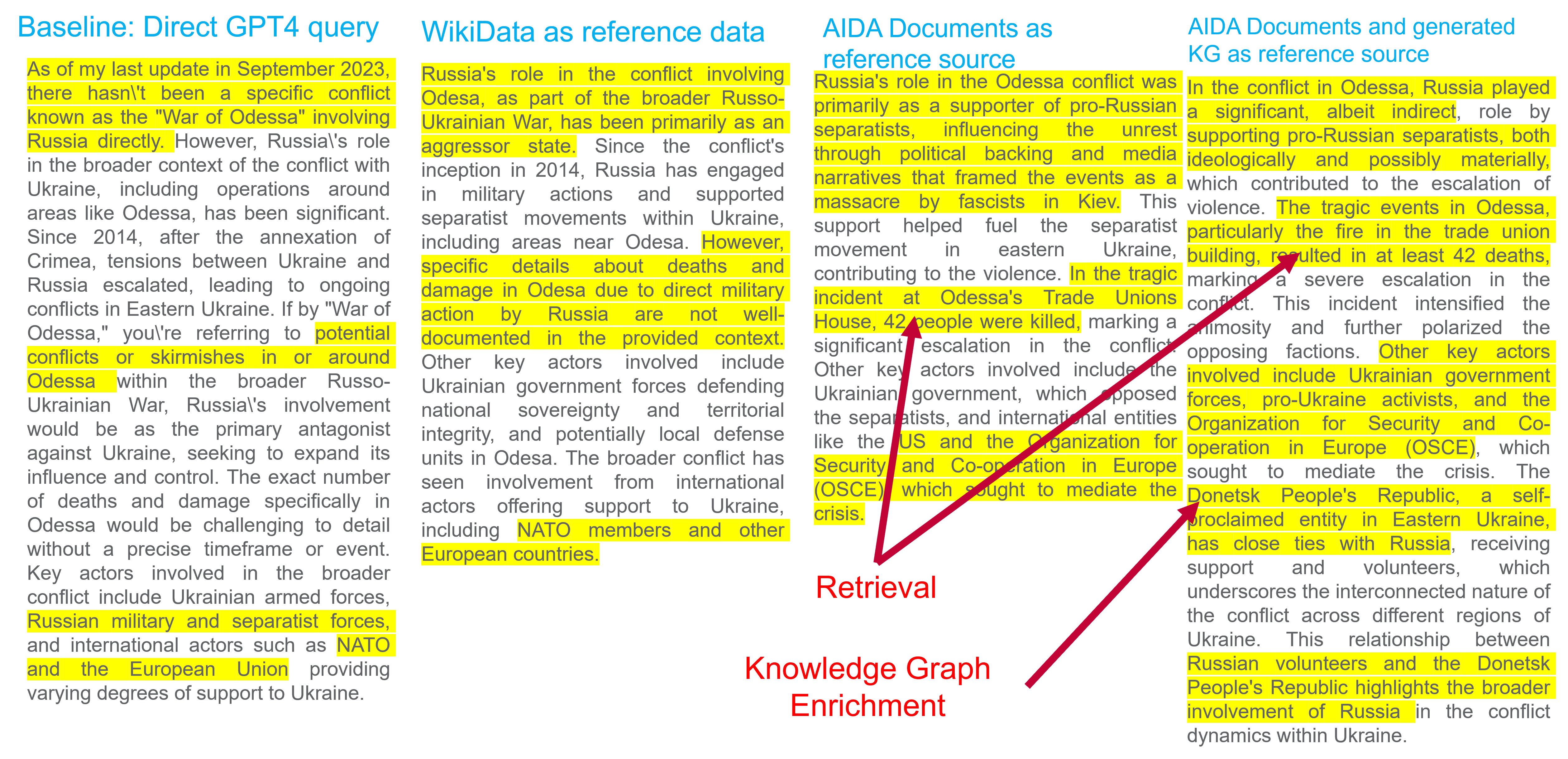}
    \captionsetup{width=0.9\textwidth}
    \caption{GraphAide results for a user query \textquoteleft Describe the role of Russia in the of war of Odessa in 100 words and list how many deaths and damage happened there. Also list other key actors involved.\textquoteright}
    \label{fig:ukr14queryres}
\end{figure*}

In the query phase, we used GraphAide to analyze the input data from multiple sources and provide responses to user queries that rely upon the curated knowledge, providing contextual information to the LLM. To demonstrate GraphAide's modularity, we set up a pipeline to use different levels of augmentation in responding to user queries. In addition to using OpenAI's GPT-4o as the baseline LLM, we also utilized WikiData, a domain-specific dataset, and an enriched KG from the domain-specific dataset to answer user queries. For the qualitative evaluation of the generated responses, we compared outputs from different augmentation levels. As shown in Figure \ref{fig:ukr14queryres}, the specificity of the generated response increases as we enrich the \textit{context} shared with LLMs. Figure \ref{fig:ukr14queryres} shows that a direct query to the LLM endpoint may fail to provide a relevant response either because the LLM is not trained on data about the specific fact or because the fact is more recent than the LLM’s cutoff date. Using a reference KB may improve the response as it grounds the user query into a set of disambiguated entities. Additionally, the use of domain-specific ground-truth documents enables sharing the provenance of the generated response. As shown in Figure \ref{fig:ukr14queryres}, the AIDA document corpus contains ground-truth related to the query and allows the LLM to provide a highly relevant response. Taking this a step further, the KG-based retrieval improves \textit{context relevancy} by including entities and concepts spread across documents in the corpus. Entities such as ``Donetsk People's Republic'' and ``Russian Volunteer'' are not necessarily mentioned in the same document in the vector database but are rather connected through a subgraph. The context generated from the subgraph is crucial for the LLM to generate additional evidence and explanations in the response. In addition to the qualitative evaluation based on user feedback, we have also identified individual components of GraphAide suitable for quantitative evaluation to measure the impact of the domain ontology and reference knowledge base. Future work will leverage quantitative metrics such as precision, recall, and retrieval relevancy to measure improvements in entity disambiguation, semantic type assignment, and KG generation tasks for a KG-based hybrid RAG.

\section{Conclusion}
Large Language Models (LLMs) offer exciting opportunities to develop digital assistents for domain-specific tasks. Advanced Retrieval Augmentation Generation (RAG) addresses explainability, scalability, and usability concerns in using LLMs and increases user confidence in the technology adoption. We prsent GraphAide, that leverages knowledge graph and subgraph matching capabilities to improve the accuracy of RAG-based applications. We present the reference architecture and its key components. We also demonstrate an instance of GraphAide to generate a Knowledge Graph (KG) from news articles describing historical ukraine-russian political conflict scenario. Future work will focus on formal evaluation to report quantitative improvements in metrics such as accuracy and relevancy.

\section*{Acknowledgement}
The research described in this paper is partially supported by the Resilience Through Data Driven, Intelligently Designed Control (RD2C) Initiative at Pacific Northwest National Laboratory (PNNL) and the United States federal government. Pacific Northwest National Laboratory is a multiprogram
national laboratory operated for the US Department of Energy (DOE) by Battelle Memorial Institute under Contract No. DE-AC05-76RL01830.

\bibliographystyle{IEEEtranN}
\bibliography{references}
\end{document}